\documentclass[reprint, amsmath, amssymb, superscriptaddress, aps]{revtex4-2}
\usepackage{color}
\usepackage{verbatim}
\usepackage{xcolor}
\usepackage{lipsum}
\usepackage{hyphenat}
\usepackage{graphicx} 
\usepackage{dcolumn} 
\usepackage{bm} 
\usepackage{tabularx}
\usepackage{hyperref} 
\usepackage{mathtools}
\usepackage{etoolbox}
\usepackage{colortbl}
\usepackage{xcolor}
\usepackage{soul}
\usepackage{float}
\usepackage{diagbox}
\usepackage[caption=false]{subfig}
\usepackage[normalem]{ulem}
\preprint{APS/123-QED}
\newsavebox{\measurebox}
\newcolumntype{l}{X}
\newcolumntype{s}{>{\hsize=0.3\hsize}X}
\newcolumntype{m}{>{\hsize=0.85\hsize}X}

\begin{document}

\title{Observation of a phase transition in KTaO$_3$ induced by residual niobium impurities}

\author{Zijun C. Zhao}\email{cindy.zhao@uwa.edu.au}
\affiliation{Quantum Technologies and Dark Matter Labs, Department of Physics, University of Western Australia, 35 Stirling Highway, Crawley, WA 6009, Australia.}
\author{Jeremy F. Bourhill}
\affiliation{Quantum Technologies and Dark Matter Labs, Department of Physics, University of Western Australia, 35 Stirling Highway, Crawley, WA 6009, Australia.}
\author{Maxim Goryachev}
\affiliation{Quantum Technologies and Dark Matter Labs, Department of Physics, University of Western Australia, 35 Stirling Highway, Crawley, WA 6009, Australia.}
\author{Aleksey Sadekov}
\affiliation{Centre for Microscopy, Characterisation and Analysis, The University of Western Australia, 35 Stirling Highway, Perth, WA 6009, Australia}
\author{Michael E. Tobar}\email{michael.tobar@uwa.edu.au}
\affiliation{Quantum Technologies and Dark Matter Labs, Department of Physics, University of Western Australia, 35 Stirling Highway, Crawley, WA 6009, Australia.}

\begin{abstract}
We report the observation of a phase transition in a KTaO$_3$ crystal, corresponding to a paraelectric-to-ferroelectric transition. The crystal was placed inside a copper cavity to form a dielectric-loaded microwave cavity, and the transition was observed to occur near 134 K. As the cavity was cooled, the frequencies of both transverse electric and transverse magnetic resonant modes decreased (corresponding to an increase in permittivity). The mode frequencies converge at the transition temperature (near 134 K) and, below this point, reverse their tuning direction, increasing their frequency with decreasing temperature. This behaviour corresponds to a decrease in dielectric permittivity and is atypical for pure KTaO$_3$. To investigate further, we conducted impurity analysis using Laser Ablation inductively coupled mass spectrometry (LA-ICPMS), revealing a significant concentration ($\sim$ 7\%) of niobium (Nb) in the crystal. This suggests that the observed phase transition is driven by residual Nb impurities, which induce ferroelectricity in an otherwise paraelectric host. Similar crystals with a lower concentration ($<$ 2\%) did not undergo a phase transition but exhibited a loss peak at this temperature. These findings have practical implications for the design of tunable devices, for example, resonator-based dark matter detectors, where low-loss material phase stability and tunability are crucial.
\end{abstract}


\maketitle

\section{Introduction}\label{sec1}

Perovskite materials such as potassium tantalate (KTaO$_3$, or KTO) and strontium titanate (SrTiO$_3$, or STO) have garnered considerable attention in both condensed matter physics and material science \cite{PhysRevLett.12.474, gastiasoro2020superconductivity, pai2018physics, PhysRevMaterials.8.124404, science.aba5511}. Both materials exhibit unique dielectric properties, particularly at low temperatures, and offer exciting possibilities for a range of applications \cite{1976888, liu2023tunable, gupta2022ktao3, sakudo1971dielectric, geyer2005microwave}. The intriguing behavior of these materials as incipient ferroelectrics, where the materials show a strong propensity for ferroelectricity, yet do not undergo a ferroelectric phase transition at ambient conditions, makes them highly interesting for both fundamental research and practical applications \cite{muller1979srti, sen2024multifunctional, kvyatkovskii2001quantum, akbarzadeh2004atomistic, zhang2023spontaneous, najev2025electronic}. These materials have been widely investigated for high permittivity ($\epsilon\propto10^2-10^3$), and low loss at low temperatures, making them ideal candidates for precision applications such as dark matter detection and other quantum technologies \cite{matsubara2016observation, caviglia2010two, barthelemy2021quasi, wadehra2020planar, kumar2021observation, massarotti2020high, choi2015quantum}. While STO has been a subject of considerable research, especially in quantum paraelectricity\cite{li2019terahertz, fujishita2016quantum, shin2021quantum, hayward2005interaction, kustov2020domain, shin2022simulating}, KTO has emerged as a more promising material for tuning devices. Although STO exhibits an interesting phase transition that has been validated in recent literature \cite{zhao2021precision}, it suffers from higher loss compared to KTO, making KTO a more favorable choice for precise tuning in this context.

One of the most notable applications of KTO lies in the development of tunable resonant cavities for dark matter detection, particularly in the search for axions. Axions, a candidate for dark matter, have yet to be directly detected, and their mass is currently unknown \cite{chadha2022axion, peccei1977cp, weinberg1978new, di2020landscape}. Resonant cavity-based detectors such as haloscopes are at the forefront of axion detection \cite{semertzidis2022axion, sikivie1983experimental, panfilis1987limits, hagmann1990results, hagmann1998results, brubaker2017first,PhysRevD.111.095007,PhysRevLett.132.031601,Quiskamp22}. However, current designs predominantly rely on mechanical tuning, which has several limitations in terms of precision and reliability, particularly at cryogenic temperatures. KTO, with its high permittivity and potential for electrical tuning, a result of ferroelectric soft mode hardening \cite{Skoromets16}, presents an excellent alternative. Compared to STO, KTO exhibits the potential for earlier onset of ferroelectricity triggered
by dopants or external conditions such as pressure, which could have significant implications for extending the temperature range over which ferroelectricity occurs. In fact, previous studies, such as that by Euclid Techlabs in 2018 \cite{garcia2023development}, proposed the use of ferroelectric KTO as a tuning mechanism for dark matter axion searches, specifically in the RADES and ADMX experiments. Recent work \cite{garcia2023development} has also demonstrated the viability of KTO thin films in dark matter axion detectors, highlighting the importance of understanding their dielectric properties for the next generation of detectors.

In this work, we report the observation of an unusual phase transition in niobium-doped KTO crystals, identified through microwave spectroscopy measurements conducted in a dielectric-loaded resonant cavity. A distinct phase transition was observed near 134 K, characterised by the convergence and subsequent divergence of the resonant frequencies, a behaviour not seen in undoped KTO. Subsequent compositional analysis revealed a significant Nb impurity concentration of approximately 7\% in the sample exhibiting the transition. In contrast, other samples with Nb concentrations below 2\% showed no such anomaly. These findings suggest that the phase transition is driven by the elevated Nb content, likely inducing a transformation from a paraelectric to a ferroelectric phase, a phenomenon absent in stoichiometric KTO. This result provides new insight into impurity-driven phase behaviour in perovskite materials and highlights the critical role of Nb doping in modifying the dielectric response of KTO.

\section{Methods}

\begin{figure}[t]
\centering
\includegraphics[width=0.75\columnwidth]{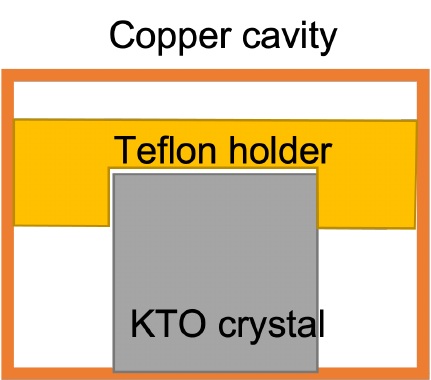}
\caption{Schematic of one of the measurement cavity setups. The cavity is constructed from oxygen-free copper (OFC) and designed to host a cylindrical KTO crystal with dimensions of 3 mm in diameter and 3 mm in height. The crystal rests within the cavity and is held securely in place by a Teflon spacer, which ensures it is held in place during both cool-down and warm-up measurement cycles. This cavity was used to measure the frequency-temperature response of sample 1, due to the loss induced by Nb impurities, the cavity walls did not contribute significantly to the loss.}
\label{fig_setup_old}
\end{figure}
\begin{figure}[t]
\centering
\includegraphics[width=1.0\columnwidth]{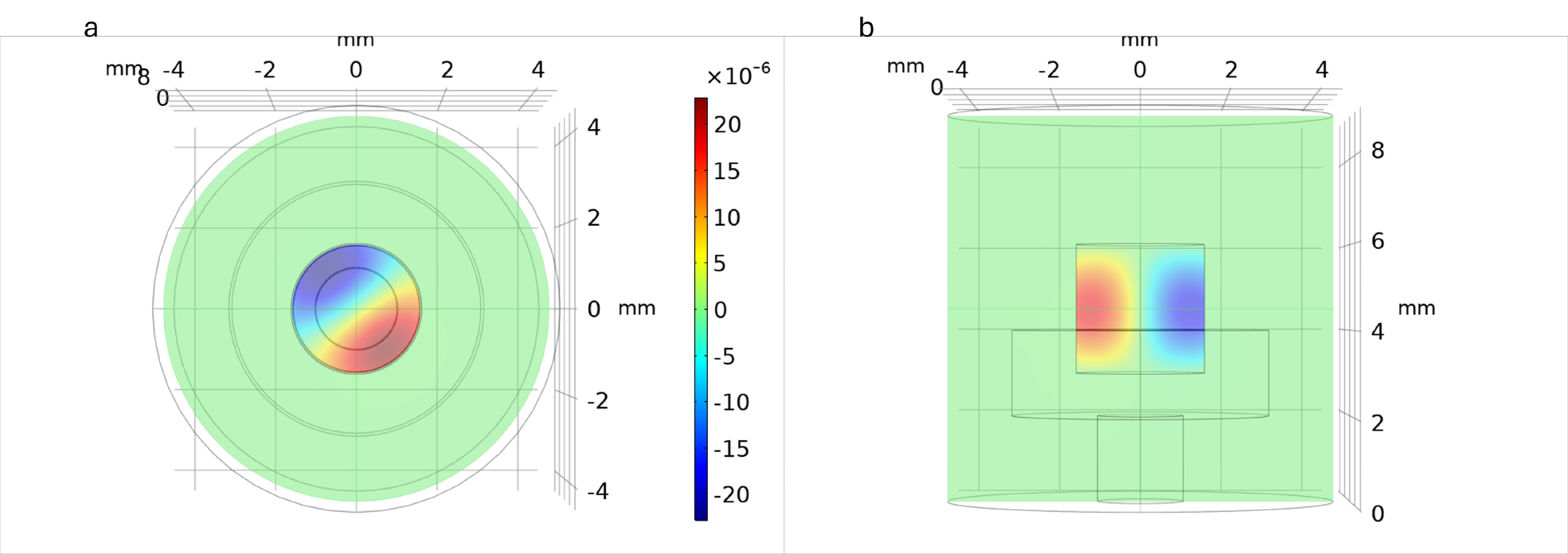}
\caption{This figure shows the electric displacement field z component $D_z$ of TM$_{1,1,1}$ mode in COMSOL simulation on a and b, b shows the picture of the KTO crystal inside the resonant cavity. The cavity is made by OFC (oxygen-free copper). The cylindrical KTO sample (3 mm diameter by 3 mm height) sits on top of a low-loss sapphire support. This cavity is used to measure the permittivity and loss tangent of sample 3 with insignificant loss contributed by the cavity surface resistance.}
\label{fig1_combine}
\end{figure}

To study the dielectric response of KTO as a function of temperature ($T$), several cylindrical single crystals, each 3 mm in both height and diameter, were sourced from three different suppliers, labeled as sample~1 (SurfaceNet), sample~2 (ALB Materials Inc.), and sample~3 (AWI Industries). Each crystal was measured individually using the same cylindrically symmetric oxygen-free copper cavity, which had an internal diameter of 8\,mm and a height of 6.5\,mm. A Teflon spacer was used to secure the crystal in place. A schematic of the measurement setup is shown in Fig.~\ref{fig_setup_old}.

Microwave radiation was coupled into the cavity using loop probes oriented at 45$^\circ$ to the cavity axis, enabling the excitation of multiple mode polarizations. The transmission spectrum $S_{21}$ was measured using a vector network analyzer (VNA) at both room and cryogenic temperatures. The cavity was cooled to 4 K using a pulse tube refrigerator, and $S_{21}$ was recorded as a function of temperature during a slow, controlled warm-up under thermal equilibrium. Modes up to 10\, GHz were monitored.

Sample~1 exhibited a prominent loss peak and an anomalous frequency shift, which were not observed in samples~2 or~3. The higher loss rates of this sample in general compared to samples 2 and 3 meant that the cavity described in Fig.~\ref{fig_setup_old} was sufficient for its initial characterization, in which the crystal makes direct contact with the cavity metallic walls. To more accurately measure the intrinsic $Q$ factors of the other samples, they are instead supported by a sapphire sample holder, shown in Fig.~\ref{fig1_combine}. The sapphire ensures no direct contact with metallic boundaries whilst maintaining good thermal conduction to the crystal. This cavity has an internal diameter of 9 mm and a height of 9 mm.

To support experimental mode identification, finite element simulations were performed using \textsc{COMSOL} Multiphysics. These simulations provided the expected frequencies of transverse magnetic (TM) and transverse electric (TE) modes, which were compared with the measured spectra to determine mode assignments.

To understand the difference between the three samples, geochemical analyses (impurity analyses) of the individual crystals were obtained by using Laser Ablation Inductively Coupled Plasma Mass Spectrometry (LA-ICPMS). A Sector Field ICP-MS (Thermo, Element XR) was used in combination with a G2 Analyte (Teledyne) laser ablation system at the Centre for Microscopy, Characterisation, and Analysis at the University of Western Australia. Both the mass spectrometer and laser system were optimized for high spatial resolution using an aperture slit of 75 x 75 $\mu$m (spot size), 4 Hz, and 1.2 j/cm$^2$ fluency. ICP-MS was operated at medium resolution to avoid potential mass interference. Initially, each crystal was scanned in {`}exploratory mode{'} to estimate the relative concentration of different elements in the samples. In this mode, we did not quantify the concentration but estimated the relative signals for most of the elements after a 5-second integration of the laser ablation signal. Using this approach, we found that elements Nb had sufficient concentrations to be relevant in explaining the onset of the phase transition around 134 K. To quantify the concentrations of these elements, we then conducted three spot analyses at three different locations on each studied crystal. Standard laser ablation data processing was employed, including background subtraction and standards quantification using US National Institute of Standards and Technology (NIST) 614 standards and NIST612. Based on the reproducibility of NIST standards, the average analytical error is around 4\%, expressed as 2SE. 

\section{Results and discussions}\label{sec2} 

\begin{figure}[t]
\centering
\includegraphics[width=1.0\columnwidth]{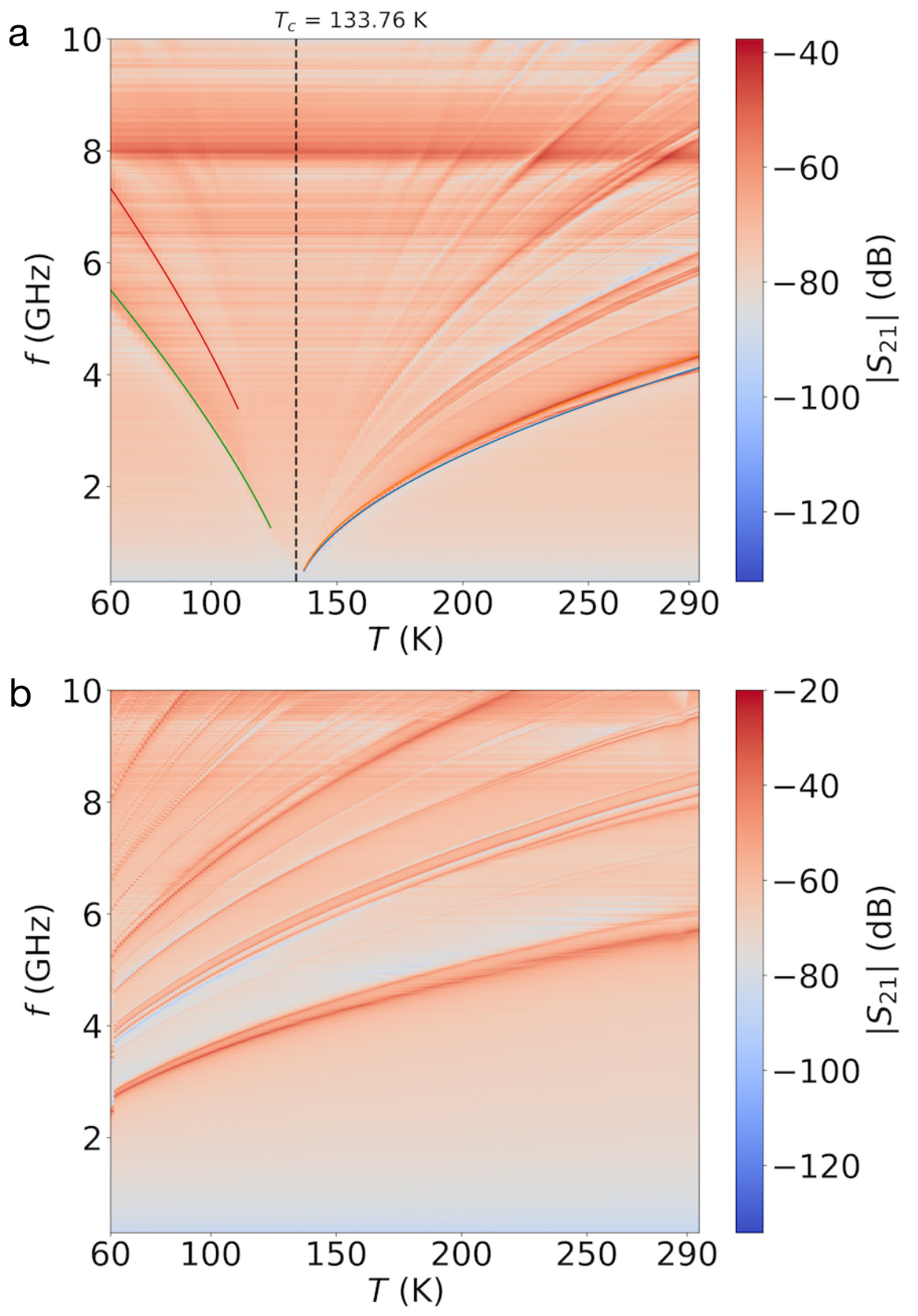}
\caption{Transmission as a function of temperature for different KTO samples. The resonant modes evolve as each sample gradually warms up, Sample 1 and sample 3 were both measured in the cavity of Fig. \ref{fig_setup_old}. a) For Sample 1, as the temperature (T) decreases, the mode frequencies shift downward and converge near 134 K. Below this temperature, the modes diverge and their frequencies begin to increase. b) For Sample 3, the mode frequencies also decrease with temperature but do not exhibit a point of convergence.}
\label{fig1}
\end{figure}

\begin{figure}[t]
\centering
\includegraphics[width=1.0\columnwidth]{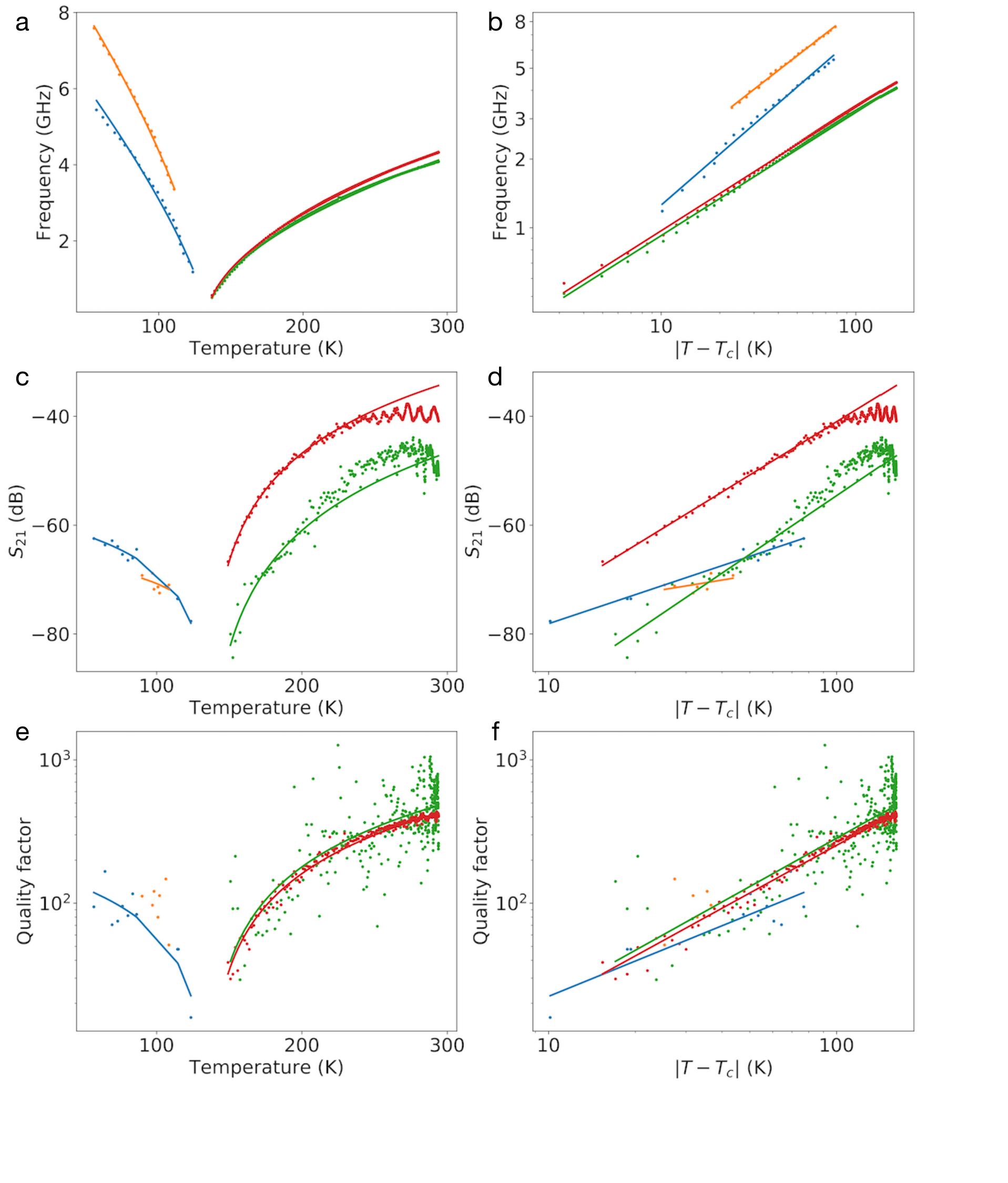}
\caption{Power law fit for various parameters vs temperature of the form $y =  a_i\cdot\lvert T - T_c \rvert ^{\beta_i} $. Frequency (a), transmission amplitude (c), and $Q$-factor (e) vs temperature. (b), (d) and (e) show the equivalent log-log plots as a function of  $\lvert T - T_c \rvert$, with straight lines highlighting the power laws and the fit parameters summarized in Tab. \ref{tab2}. The TM$_{1, 1, 1}$ mode is plotted in green above $T_c$, and in blue below $T_c$. The TE$_{0, 1, 1}$ mode is plotted in red above $T_c$ and in orange below $T_c$.}
\label{fig3}
\end{figure}

\begin{table}[h]
\begin{center}
\begin{minipage}{1\columnwidth}
\caption{Fitted power law coefficient, $\beta_i$, for various parameters plotted in Fig .\ref{fig3}.}
\label{tab2}
\begin{tabular}{|c|c|c|c|c|}
\hline
\shortstack{power law \\ coefficient \\ vs \\ resonant mode}& \shortstack{1(green) \\ TM$_{1,1,1}$\\above \\ $T_c$}  & \shortstack{2(red) \\TE$_{0,1,1}$ \\above \\ $T_c$} & \shortstack{3(blue) \\ TM$_{1,1,1}$\\below \\$T_c$} & \shortstack{4(orange) \\ TE$_{0,1,1}$ \\ below \\ $T_c$} \\
\hline
$\beta_f$ & \shortstack{0.540 \\$\pm$ 0.001} & \shortstack{0.5411 \\$\pm$ 0.0009} & \shortstack{-0.74 \\$\pm$ 0.02} & \shortstack{-0.663 \\$\pm$ 0.007} \\
\hline
$\beta_{S_{21}}$ & \shortstack{1.79 \\$\pm$ 0.09} & \shortstack{1.62 \\$\pm$ 0.02} & \shortstack{-0.83 \\$\pm$ 0.07} & \shortstack{-0.5 \\$\pm$ 0.1} \\
\hline
$\beta_Q$  & \shortstack{1.12 \\$\pm$ 0.06} & \shortstack{1.104 \\$\pm$ 0.008} & \shortstack{-0.6 \\$\pm$ 0.1} & \\
\hline
$\beta_\varepsilon$ & \shortstack{-1.082 \\$\pm$ 0.002} & & \shortstack{-1.48 \\$\pm$0.03} & \\
\hline
\end{tabular}
\end{minipage}
\end{center}
\end{table}

Fig. \ref{fig1} shows the transmission parameter, $S_{21}$, as a function of frequency and temperature during a gradual warm-up of the cavity. Data for sample 1 (Fig. \ref{fig1}a) and sample 3 (Fig. \ref{fig1}b) are shown for comparison. For sample 1, as the temperature decreases, the frequencies of the resonant modes decrease and converge around a transition temperature $T_c$. Below $T_c$ the frequencies increase again with further cooling, indicating anomalous behaviour. In contrast, sample 3 shows a monotonic decrease in resonant frequencies with temperature, but without convergence near a critical point. Sample 2 (not shown) exhibits behaviour similar to Sample 3. These observations suggest that samples 2 and 3 remain paraelectric, while sample 1 likely undergoes a phase transition from a paraelectric to a ferroelectric state near $T_c$. To further investigate this behaviour, a power law fit was applied to the resonant frequencies versus $\lvert T - T_c \rvert$. The best global fit across four modes yields a transition temperature of $T_c=133.76  \pm  0.01$ K. The fitted powers from this analysis are summarised in Tab. \ref{tab2}.

The temperature dependence of the resonant frequency $f$ for two lower-order modes (TE$_{0,1,1}$ and TM$_{1,1,1}$) in Sample 1 is shown in the Fig. \ref{fig3}a. Solid lines indicate power-law fits of $f$ versus $\lvert T-T_c \rvert$ in the form of $f = a_f \cdot \lvert T-T_c \rvert ^{\beta_f}$, with the corresponding exponent $\beta_f$ listed in Tab. \ref{tab2}. Similarly, the variation of $S_{21}$ with temperature is shown in the Fig. \ref{fig3}c, along with power-law fits in the form $S_{21} = a_{S_{21}} \cdot \lvert T-T_c \rvert ^{\beta_{S_{21}}}$ using exponent $\beta_{S_{21}}$ also reported in Tab. \ref{tab2}. The temperature dependence of the quality factors Q is shown in the Fig. \ref{fig3}e, fitted with power laws in the form of $Q = a_Q \cdot \lvert T-T_c \rvert ^{\beta_Q}$ with the corresponding exponent $\beta_Q$ listed in Tab. \ref{tab2}.  These fitted power-law behaviors consistently point to a phase transition near $T_c=134 K$.

To investigate the source of the phase transition around $T_c$ in sample 1, we performed impurity analyses of three samples using LA-ICPMS and found that there is more than 7\% Nb concentration in sample 1, while Nb concentration is low in sample 2 and sample 3. The details of Nb concentration and source of the samples are summarized in the Table. \ref{tab3}. Thus, the phase transition can be attributed as Nb-induced phase transition. The results in this paper show that only when the Nb concentration is larger than a threshold, the phase transition occurs. If the Nb concentration is below the threshold, the KTO remains as an incipient ferroelectric material when cooled down to a low temperature. Nb impurities can induce ferroelectricity in KTO by disrupting the crystal's centrosymmetric structure, leading to off-center displacements of Nb ions. These displacements create local dipoles within the lattice, promoting a spontaneous polarization characteristic of ferroelectric materials. Prior work indicates that Nb doping lowers the threshold for ferroelectric phase transitions in KTO \cite{hochli1977quantum} observing phase transitions with Nb concentrations as low as 0.5\% . Additionally, computational models \cite{eglitis1998semi} suggest that Nb impurities favourably occupy off-center positions in the lattice, further supporting the induction of ferroelectricity through such structural distortions.

\begin{table}[ht]
\begin{center}
\begin{minipage}{1\columnwidth}
\caption{Impurities concentration for different samples in the cavity of Fig. \ref{fig1_combine} }
\label{tab3}%
\begin{tabular}{|c|c|c|c|}
\hline
sample number & 1 & 2 & 3 \\
\hline
source & SurfaceNet & ALB material & AWI industries \\
\hline
Nb concentration \% & 7.1 $\pm$ 0.3 & 1.84 $\pm$ 0.07 & 0.102 $\pm$ 0.004 \\
\hline
\shortstack{room temperature \\quality factor} & 1.84$\times10^3$ & 2.76$\times10^3$ & 2.95$\times10^3$ \\
\hline
\end{tabular}
\end{minipage}
\end{center}
\end{table}

\begin{figure}[t]
\centering
\includegraphics[width=1.0\columnwidth]{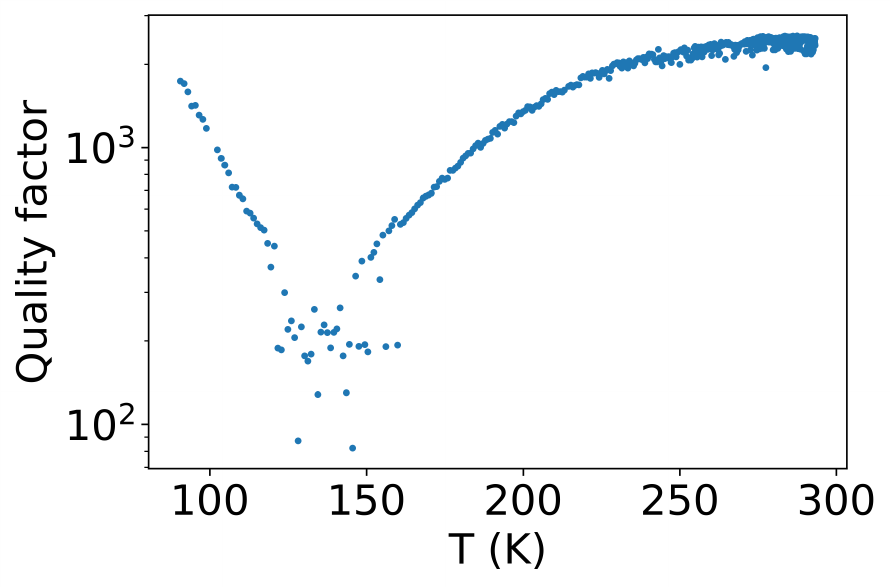}
\caption{Quality factor vs Temperature T for sample 3 in cavity shown in Fig. \ref{fig1_combine}. }
\label{fig6}
\end{figure}

Quality factor Q vs temperature for sample 3 in cavity of Fig .\ref{fig1_combine} is shown in Fig. \ref{fig6}. Although there is no phase transition in sample 3 as the Nb concentration is low, we can see a drop in Q at around the transition temperature $T_c$. We suspect there might be a small dielectric island that goes through the phase transition rather than the whole crystal, which adds loss at $T_c$. This result shows that even a small amount of Nb impurities can have an effect on the behavior of the KTO crystal.

\begin{figure}[ht]%
\centering
\includegraphics[width=1.0\columnwidth]{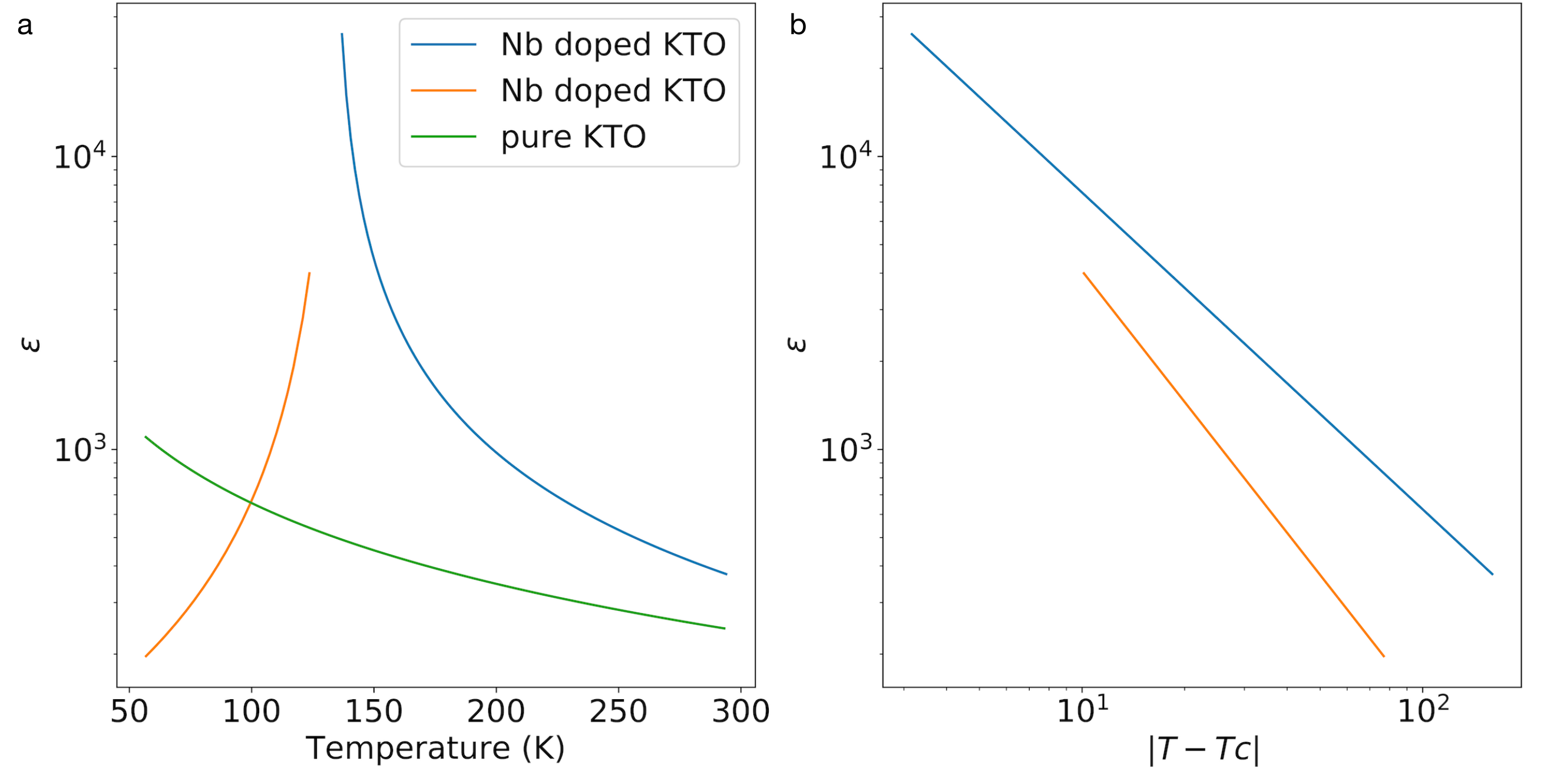}
\caption{a) Permittivity $\varepsilon$ versus temperature of samples 1 and 3. b) Fits for the logarithm of the permittivity versus the logarithm of temperature $\mid T-Tc \mid$ is shown, where $Tc$ is the transition temperature. }
\label{fig5}
\end{figure}

To characterize dielectric constant of the KTO, we compare experimental frequency vs temperature data and simulated frequency vs permittivity from COMSOL, and thus calculate the permittivity vs temperature. The permittivity $\varepsilon$ vs temperature for Nb doped sample 1 and pure sample 3 is plotted in Fig. \ref{fig5}a, and logarithm of permittivity vs $\log \lvert T - T_c \rvert$ is shown in Fig. \ref{fig5}b and slopes are summarized in Table. \ref{tab2}. Note that to characterize material property of pure sample better, we use the cavity in Fig. \ref{fig1_combine}, where KTO sits in a sapphire post to minimize the loss while maintain good thermal contact. We can confirm the permittivity of pure sample (sample 3) at room temperature is 239 $\pm$ 8 and similar to what was measured in the literature \cite{geyer2005microwave}.

\section{Conclusion}

In this paper, we analysed three KTO samples and observed an unexpected phase transition near 134 K in one of the samples. Impurity analysis via laser ablation mass spectrometry revealed that it is likely attributed to a niobium (Nb)-induced phase transition. The sample exhibiting this transition contained approximately a 7\% concentration of Nb, whereas the other samples, with below 2\% concentration, did not show such behaviour. The sample with the lowest Nb concentration (sample 3) exhibited the lowest loss tangent. This finding shows the importance of low-impurity concentration to obtain good dielectric properties for KTO, necessary to engineer low-loss tuneable devices, for example, for applications in dark matter detection and precision measurement.

\bibliography{KTO}
\bibliographystyle{unsrt}


\end{document}